\documentclass[%
  anonymous=false,%
  authordraft=false,%
  format=sigconf,%
  review=false,%
  screen=false,%
  timestamp=true,%
]{acmart}

\copyrightyear{2022}
\acmYear{2022}
\setcopyright{acmlicensed}
\acmConference[ICSE '22 Companion]{44th International Conference on Software Engineering Companion}{May 21--29, 2022}{Pittsburgh, PA, USA}
\acmBooktitle{44th International Conference on Software Engineering Companion (ICSE '22 Companion), May 21--29, 2022, Pittsburgh, PA, USA}
\acmPrice{15.00}
\acmDOI{10.1145/3510454.3516862}
\acmISBN{978-1-4503-9223-5/22/05}

\usepackage{defs}
\usepackage{todonotes}
\usepackage{listings}
\usepackage{float}
\usepackage{cleveref}
\graphicspath{{./}{./img/}}

\newcommand{\Gamekins}{\textsc{Gamekins}\xspace}
\newcommand{\Jenkins}{\textsc{Jenkins}\xspace}
\newcommand{\SonarLint}{\textsc{SonarLint}\xspace}

\begin{document}

\title{Gamekins: Gamifying Software Testing in Jenkins}

\author{Philipp Straubinger}
\orcid{0000-0002-9265-5789}
\email{philipp.straubinger@uni-passau.de}
\affiliation{%
  \institution{University of Passau}
  \city{Passau}
  \country{Germany}
}

\author{Gordon Fraser}
\orcid{0000-0002-4364-6595}
\email{gordon.fraser@uni-passau.de}
\affiliation{%
  \institution{University of Passau}
  \city{Passau}
  \country{Germany}
}

\renewcommand{\shortauthors}{Straubinger and Fraser}



\begin{abstract}
  Developers have to write thorough tests for their software in order to find
  bugs and to prevent regressions. Writing tests, however, is not every
  developer's favourite occupation, and if a lack of motivation leads to a lack
  of tests, then this may have dire consequences, such as programs with poor
  quality or even project failures.
  This paper introduces \Gamekins, a tool that uses gamification to motivate
  developers to write more and better tests. \Gamekins is integrated into the
  \Jenkins continuous integration platform where game elements are based on
  commits to the source code repository: Developers can earn points for
  completing test challenges and quests posed by \Gamekins, compete with other
  developers or developer teams on a leaderboard, and are rewarded for their
  test-related achievements.
  A demo video of \Gamekins is available at
  \url{https://youtu.be/qnRWEQim12E};
  The tool, documentation, and source code are available at
  \href{http://gamekins.org}{https://gamekins.org}.
\end{abstract}


\maketitle

\section{Introduction}

Achieving high software quality is a challenge, as witnessed for example by regular reports of economic loss caused by poor software quality~\cite{krasner2021cost}. Software testing is an essential factor for achieving high software quality~\cite{DBLP:conf/esem/SantosMCSCS17}, but unfortunately, many developers are not highly motivated to include thorough testing in their everyday work. The reasons for this are manifold, and include the low recognition of software testing in their company, testing being perceived as dull and monotonous, and tests not making a visible impact on the progress of software projects~\cite{DBLP:journals/software/WeyukerOBP00,DBLP:journals/corr/WaychalC16,DBLP:journals/infsof/DeakSS16}.

Motivation in software engineering is a common problem addressed by various approaches in the last years \cite{8370133, ryan2000self}. One promising solution to motivate developers is the concept of \emph{gamification}, which generally consists of adding game elements to non-game related tools and contexts. These game elements can be everything that is characteristic for games like the competition aspects with points, teams and leaderboards~\cite{DBLP:conf/mindtrek/DeterdingDKN11}. Gamification has been successfully applied to various parts of software engineering, including software construction and requirements, as well as in software enigneering process and management \cite{DBLP:conf/icse/BarretoF21}. We introduce \Gamekins, a tool that applies gamification to software testing.

\Gamekins aims to motivate developers to write more tests by integrating
gamification in \Jenkins, one of the most popular CI environments in
industry. At the end of each build process of a software project with \Jenkins
the source code and test results are analysed, and tasks tailored to each
developer are generated. These tasks vary from writing new tests to improving
existing code, and by solving them developers improve the quality of the test
suite.
Developers are incentivised to complete these tasks using various gamification
elements, such as points for solving tasks, being rewarded for good testing
with achievements, and competing with other developers and development teams
through a leaderboard.

The integration into \Jenkins makes the gamification accessible to developers, who are likely already familiar with the usage of \Jenkins and thus require no training to use \Gamekins. This is an important point, because software projects tend to be deadline driven and developers may not have the time to explore and learn a new environment. A further advantage of implementing the gamification as a CI plugin is that source code as well as test results are readily available, and the challenges can be focused on the aspects of the code a developer is currently working on.

\section{Gamification elements}
\Gamekins provides a variety of gamification elements, in particular challenges, quests, leaderboards and achievements.

\subsection{Challenges}
\begin{figure}
	\includegraphics[width=\linewidth]{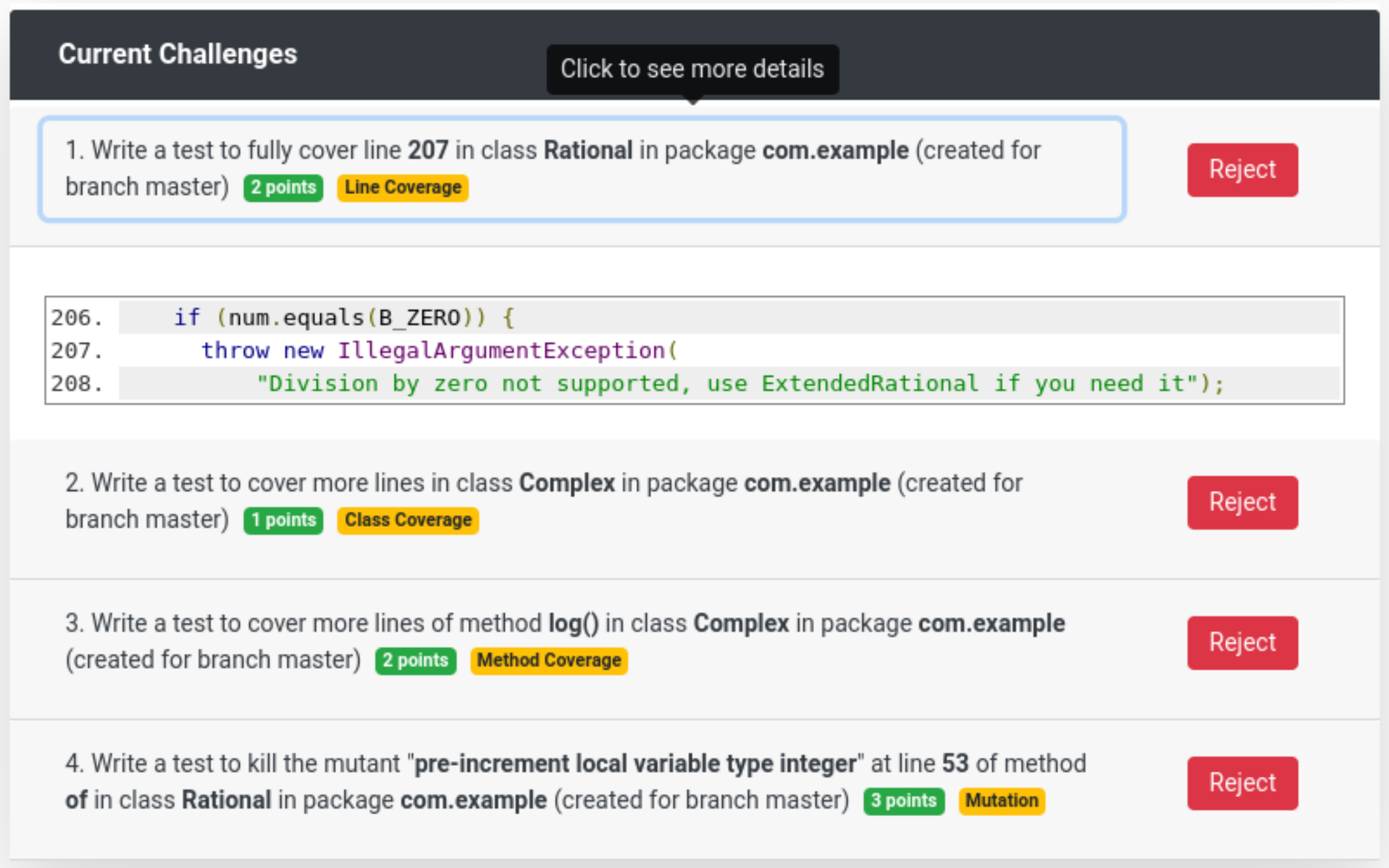}
	\caption{Overview of current challenges, showing a Line-, Method- and Class Coverage-Challenge as well as a Mutation Test Challenge. The Line Coverage Challenge is expanded with the relevant code snippet for the challenge displayed.}
	\label{fig:challenges}
\end{figure}

\Gamekins provides a total of six different types of challenges, where each challenge gives the developer a specific task to solve. The following types of challenges are supported:
\begin{itemize}
	\item \textbf{Build Challenge}: This basic challenge tasks the developer with fixing a build when it fails. Solving this challenge leads to a reward of 1 point. Currently, this kind of challenge is generated every time the build fails. We are planning to investigate alternative strategies like limiting the number of generated challenges by number or time.
	\item \textbf{Test Challenge}: This is a generic testing challenge, which tasks the developer to write at least one more test, without narrowing down which part of the code to target with that test. Solving this challenge leads to a reward of 1 point.
	\item \textbf{Class Coverage Challenge}: This challenge requires the developer to cover more lines of code in a chosen target class. The points awarded for this type of challenge depend on the current level of coverage: If the coverage exceeds a certain threshold (e.g., 80 \%) then the developer can gain 2 points, otherwise only 1 point.
	\item \textbf{Method Coverage Challenge}: This challenge tasks the
developer to improve the coverage of a target method. The points awarded depend
on the current level of coverage, with 2 points awarded if coverage on the
class is already high and 1 point awarded otherwise.
	\item \textbf{Line Coverage Challenge}: This challenge picks a specific line of code that is not fully covered, and the task is to improve coverage (i.e., either to cover the line, or to increase the branch coverage of the line). The points for a line coverage challenge again depend on the overall coverage on the class, with 3 points being awarded if coverage is already high, and otherwise 2 points.
	\item \textbf{Mutation Test Challenge}: This challenge requires the developer to  kill a specific mutant by adding or modifying a test. Since mutation challenges are the most difficult type of challenge, they are rewarded with 4 points.
	\item \textbf{Smell Challenge}: This challenge is created by analysing the target class with \SonarLint\footnote{\url{https://www.sonarlint.org/}} and choosing one of the found smells. Smells can be detected in both source and test files, and removing these code \cite{DBLP:books/daglib/0019908} or test \cite{van2001refactoring} smells leads to a reward of 1 to 4 points, based on the severity of the smell.
\end{itemize}
Clicking on a challenge in the overview (\cref{fig:challenges}) reveals additional information about the challenge, such as the code snippet mentioned in the description, as well as an explanation what to do. This is especially important for mutation challenges, where the modified source code and the original one are shown.

\subsection{Quests}
\begin{figure}
	\includegraphics[width=\linewidth]{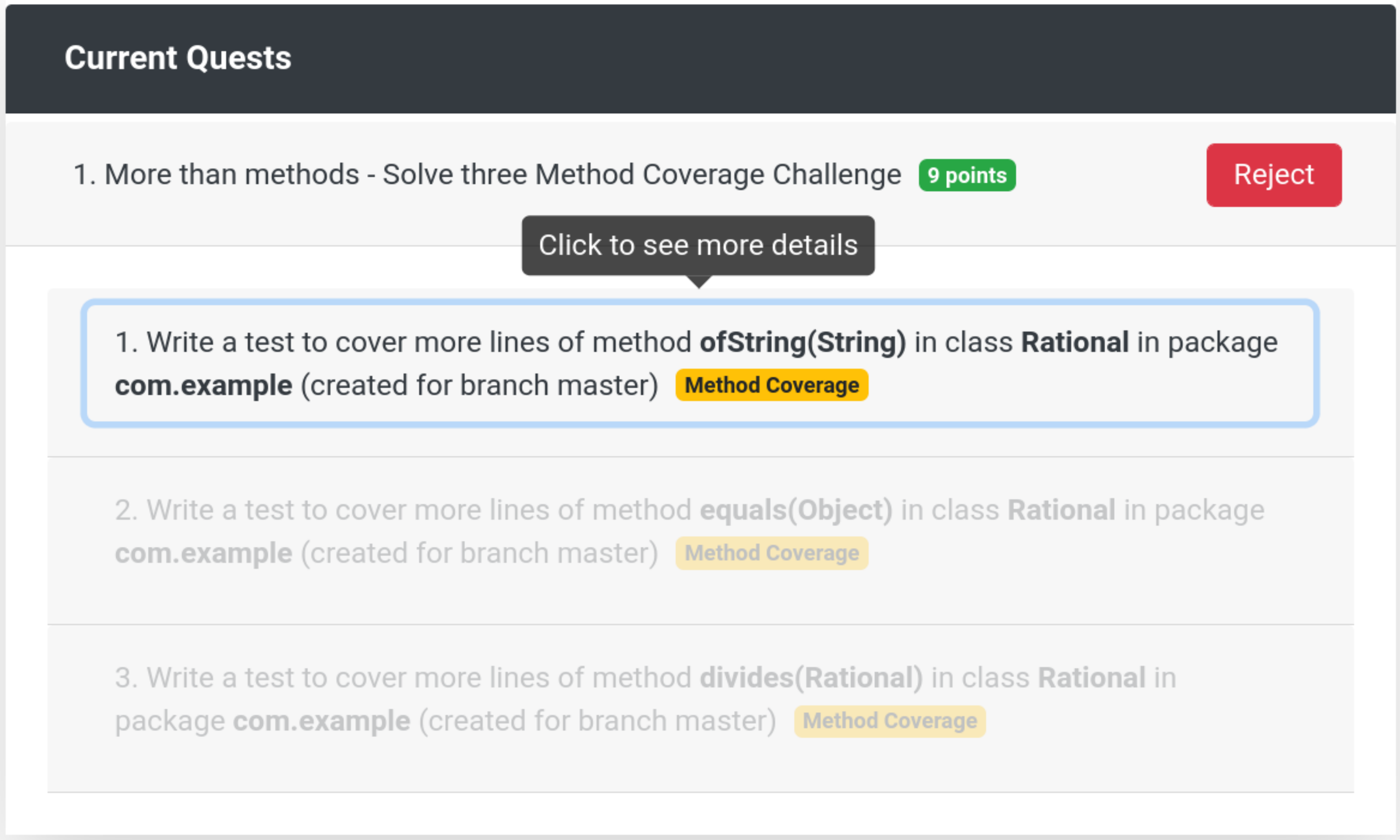}
	\caption{Example quest consisting of three Method Coverage Challenges, with the active one highlighted.}
	\label{fig:quests}
\end{figure}

Multiple challenges can be grouped into quests, which are sequences of
individual challenges that have to be solved one after the other. The current
challenge to be solved as next step in the quest is enabled and can be expanded
to reveal more details as can be seen in \cref{fig:quests}. Successor steps
are disabled and cannot be viewed until it is their turn to be solved. The
points of the whole quest, i.e., the sum of the points of the constituent
challenges and one additional point for each step, are only added to
the score of the participant if all steps are solved successively. The
following types of quests are supported by \Gamekins:

\begin{itemize}
	\item \textbf{Test Quests} consist of three test challenges.
	\item \textbf{Package Quests} consist of three different challenges in the same package.
	\item \textbf{Class Quests} consist of three Class Coverage Challenges for the same class.
	\item \textbf{Method Quests} consist of three Method Coverage Challenges for the same class.
	\item \textbf{Line Quests} consist of three Line Coverage Challenges for the same class.
	\item \textbf{Expanding Quests} consist of a Line Coverage Challenge, a Method Coverage Challenge, and a Class Coverage Challenge for the same class in this order.
	\item \textbf{Decreasing Quests} consist of a Class Coverage Challenge, a Method Coverage Challenge, and a Line Coverage Challenge for the same class in this order.
	\item \textbf{Mutation Quests} consist of three Mutation Challenges for the same class.	
	\item \textbf{Smell Quests} consist of three Smell Challenges for the same class.
\end{itemize}

\subsection{Leaderboards}
\begin{figure}
	\includegraphics[width=\linewidth]{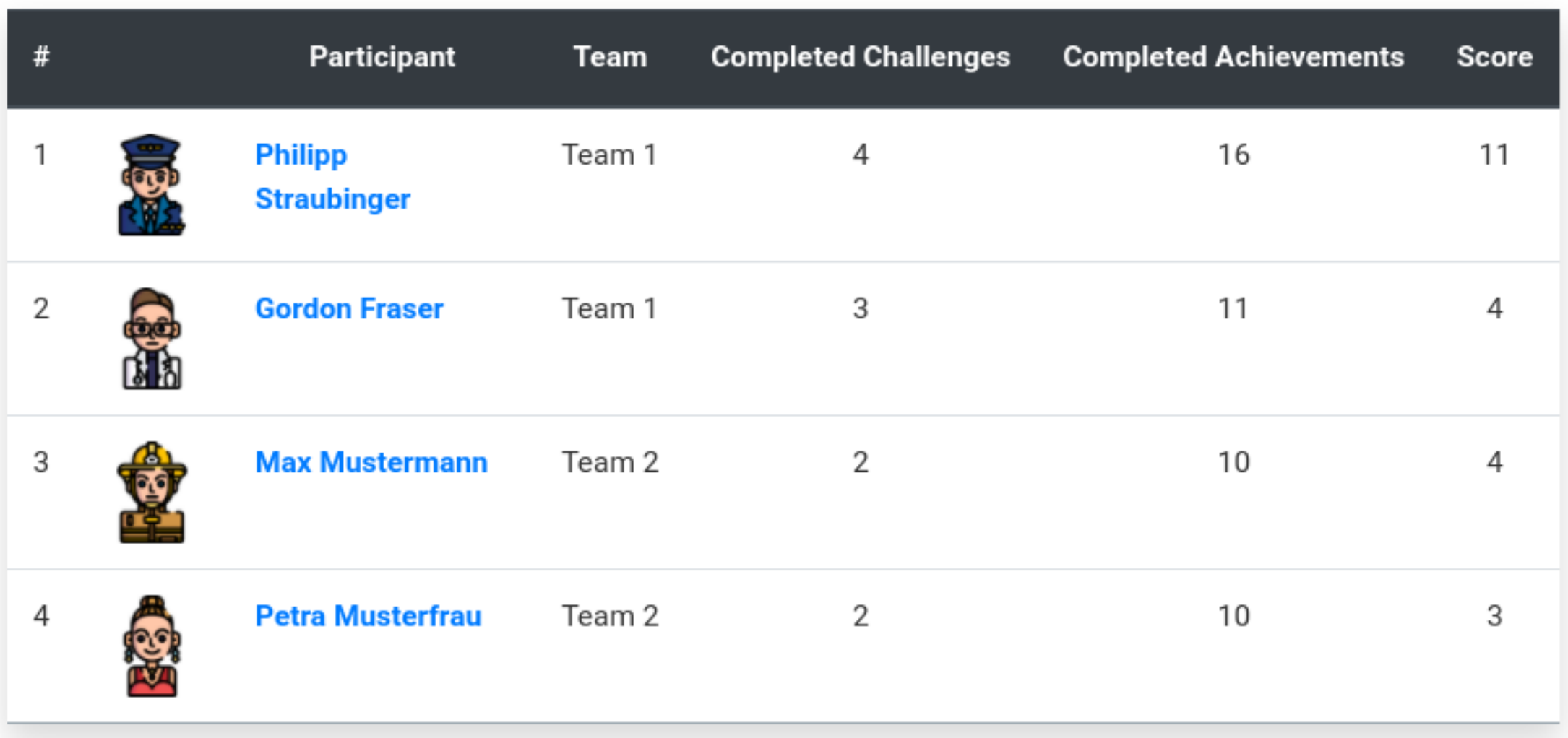}
	\caption{The leaderboard for a project shows the completed challenges and achievements as well as the score of each user. In addition, each user has its own avatar and the link on the name leads to the achievements of them.}
	\label{fig:leaderboard}
\end{figure}

By completing challenges and quests users gain points, and all users are ranked
on a project leaderboard as shown in \cref{fig:leaderboard}. Besides the
points, the leaderboard also shows the total number of completed challenges and
achievements to increase competition between developers. In addition to the
project leaderboard, a team leaderboard displays the overall points, challenges
and achievements gained by different teams.
\Jenkins allows grouping projects in folders, and \Gamekins also displays
leaderboards for folders, which enables \Gamekins to be used beyond the
confines of an individual project. Each participant can individualize their
experience by choosing one of 50 avatars for themselves to be displayed on the
leaderboard.

\subsection{Achievements}
\begin{figure}
	\includegraphics[width=0.9\linewidth]{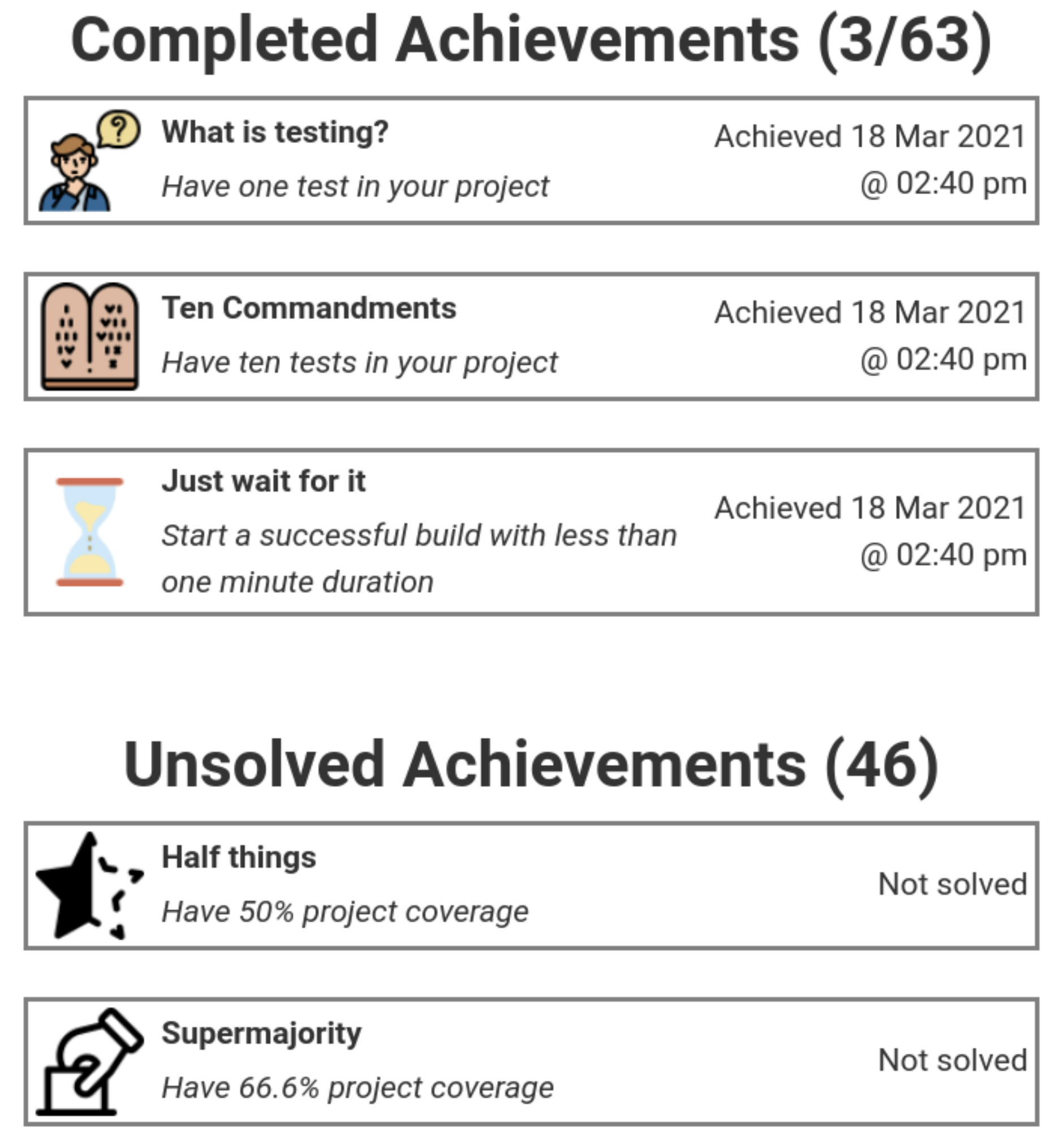}
	\caption{The list of achievements distinguishes completed, unsolved and secret achievements. Each achievement consists of a title, a description, a  date when it was solved and an icon, which is coloured after the achievement is solved.}
	\label{fig:achievements}
\end{figure}

Developers are rewarded for their test-achievements, i.e., when they exhibit
certain behavior or perform certain actions independent of specific
challenges. Achievements have incremental difficulty, ranging from easy tasks,
like having a test in a project, to difficult ones like achieving 100\%
coverage. \Gamekins currently implements 63 achievements; developers get to see
a list of obtained and open achievements (\cref{fig:achievements}), but some
achievements are secret and only revealed when they are completed. Achievements
are always solved by actions of individual participants (e.g., adding new
tests). There are two types of achievements: Individual ones, like solving a
certain number of challenges, or achievements at the project level like having
a specific coverage in the project. While project level achievements are
satisfied by a commit from an individual user, every participant in the project
receives the achievement.

\section{How Gamekins works} \label{sec:workflow}

Developers commit their code changes to a version control system, which then
triggers a job in the \Jenkins CI environment. \Jenkins executes a build of the
job together with the tests of the project, and then invokes \Gamekins.
\Gamekins takes the information of the run and the version control system to
update the challenges, quests, and achievements, as well as the leaderboard.

If a developer already has open challenges or quests, \Gamekins checks whether
these are solved and generates new ones if necessary. In addition, all
remaining unsolved achievements are checked for completion. \Gamekins also
checks whether all currently open challenges and quests are still solvable and,
for example, the specific code fragment has not been deleted.

Whenever challenges are solved, new challenges are generated such that there is
always the same (configurable) number of open challenges. When generating a new
challenge, \Gamekins considers only recently changed classes by the specific
developer according to the commit history. This has the advantages that it
takes less time and resources for generation as well that the developers are
already familiar with the class used for a challenge, since they changed
something in there recently. To generate a challenge, \Gamekins first ranks all
classes of the project under test based on the current code coverage, then
probabilistically selects a class such that those with lower coverage have a
higher chance of being selected; finally, a challenge type is randomly
selected, and a random instance for this type is generated. 

Mutation Test Challenges are based on information that \Gamekins produces using
\textsc{Moco}\footnote{\url{https://github.com/phantran/moco}}, a Maven-plugin
for mutation analysis. We developed \textsc{Moco} independently of \Gamekins
for better interchangeability and the need of the instrumented code during test
execution. \Gamekins generates candidate Mutation Test Challenges for all live
mutants reported by \textsc{Moco}.

Quests are generated by checking the prerequisites of the possible quests (e.g., if a sufficient number of required challenges can be generated) and choosing one of them randomly. The different steps or challenges are generated as they would be stand-alone, but depending on the quest type with already chosen classes.

Developers can reject challenges and quests with the indication of a reason to
further improve \Gamekins. Reasons may vary from difficulties to unsolvable
ones up to code fragments, which should not be tested. Already rejected
challenges will not be used again as challenges and the classes of rejected challenges
on class level will not be considered for new challenges either.

\section{Setting up Gamekins}
\begin{figure}
	\includegraphics[width=\linewidth]{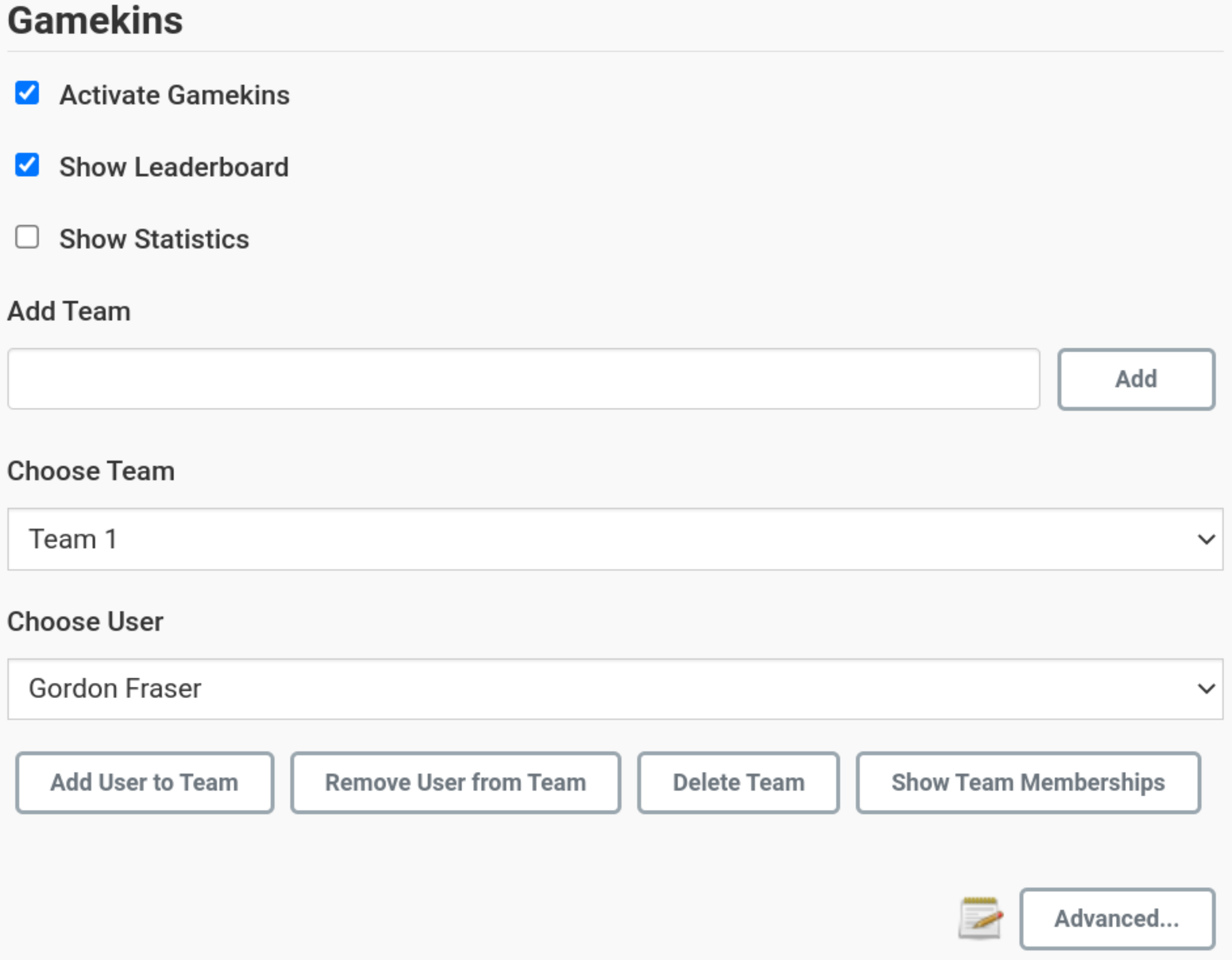}
	\caption{The Configuration section of a job in \Jenkins allows enabling \Gamekins, the leaderboard, and the statistics, as well as management of developer teams.}
	\label{fig:jobconfig}
\end{figure}

\begin{lstlisting}[float, caption=Example Jenkinsfile to configure \Gamekins., label=lst:jenkinsfile]
pipeline {
	agent any
	stages {
		stage('Test') {
			steps {
				sh 'mvn -B clean test'
			}
		}
		stage('Mutation') {
			steps {
				sh 'mvn -B m0c0:moco'
			}
		}
	}
	post {
		always {
			gamekins jacocoCSVPath: '**/target/site/jacoco/jacoco.csv', jacocoResultsPath: '**/target/site/jacoco', mocoJSONPath: '**/target/moco/mutation/moco.json', searchCommitCount: 50
		}
	}
}
\end{lstlisting}

\Gamekins can easily be added to an existing instance of \Jenkins by using the
already compiled plugin from
the git repository\footnote{\href{http://gamekins.org}{https://gamekins.org}} or by compiling it from
source. The plugin adds a new section to the configuration of different jobs in
\Jenkins with the purpose to enable \Gamekins for this particular job, as shown
in \cref{fig:jobconfig}. This is the activation point of \Gamekins and also
allows to enable different options, such as showing and hiding the leaderboard
as well as the statistics, which contains the project in an anonymized data
dump for evaluation purposes. In addition, new teams can be created and members
can be added to them. There is also a reset option, which deletes all
information collected by \Gamekins for the project. An advanced section
contains the options to adjust the number of current challenges and quests,
which are set to three and one by default. The configuration is also available
in a stripped-down version for folders in \Jenkins, which allows to group
projects together and therefore compete against each other across projects.

\Jenkins provides so called \emph{Post-build Actions} or \emph{Publishers}, which are executed after the build itself, including tests, are finished. That is where \Gamekins is triggered and therefore the publisher for \Gamekins has to be added to the configuration in order to get it to work.
This adds four additional options to the configuration with paths to specific files in the project as well as the number of commits to be searched (see \cref{sec:workflow}). This publisher is replaced by a call to \Gamekins if the project is configured as a \emph{Pipeline} in \Jenkins as shown in \cref{lst:jenkinsfile}. This so called Jenkinsfile shows the workflow of the execution of a job in \Jenkins. At first, the project itself is built and tested (Lines 4--8), with the generation of mutants afterwards (Lines 9--13). Only in the end, \Gamekins is started with the option in lines 15--19.

A new section is also added to the configuration of each user, with a text
field containing the git names, which link commits with \Jenkins users. In our
experience, the main reason why no challenges or quests are generated, are
wrong or missing git names. Users can also change their avatars here and
decide, whether they want to receive e-mail notifications from \Gamekins.
Each email includes all events that happened during the last execution of
\Gamekins, like the generation of challenges, or the completing of
achievements. Useful links are also provided to give the developer a fast and
simple opportunity to check his current progress in \Gamekins.

\section{Conclusions}
\Gamekins currently provides challenges regarding build failures, numbers of tests, coverage improvements and killing mutants. \Gamekins can be integrated with both, pipeline and non-pipeline projects, in \Jenkins for projects using Java or Kotlin with JUnit and JaCoCo. It includes various gamification elements like challenges and quests with points as rewards, achievements and leaderboards. We plan to continuously extend the challenges and other game-mechanics in \Gamekins.
A first evaluation of \Gamekins in an internal software project gave promising results: Over a period of five weeks, the coverage of the project increased by 7 \%, and the number of tests by almost 20 \%. The motivation of most participants to write tests increased during the evaluation phase. They also wrote more tests than usual and every participant found at least one bug in the project.
We invite researchers and practitioners to try out \Gamekins, and to share their experience with us:\\

\centerline{\href{http://gamekins.org}{https://gamekins.org}}

\begin{acks}
This work is supported by the DFG under grant FR 2955/2-1.
\end{acks}

\newpage

\bibliographystyle{ACM-Reference-Format}
\bibliography{related}

\end{document}